\documentclass{PoS}

\usepackage{amsmath}
\newcommand{\tauint}{\tau_{\rm int}}

\title{Investigating the critical slowing down of QCD simulations}

\ShortTitle{Investigating the critical slowing down of QCD simulations}

\author{\speaker{Stefan Schaefer}\\
        Humboldt Universit\"at zu Berlin, Institut f\"ur Physik, 
	Newtonstr. 15, 12489 Berlin, Germany\\
        E-mail: \email{sschaef@physik.hu-berlin.de}}

\author{Rainer Sommer and Francesco Virotta\\
        NIC/DESY, Platanenallee 6, 15738 Zeuthen, Germany\\
        E-mail: 
	\email{Rainer.Sommer@desy.de},
        \email{Francesco.Virotta@desy.de}
	}

\abstract{
Simulations of QCD are known to suffer from serious critical slowing down
towards the continuum limit. This is particularly prominent in the topological
charge. We investigate the severeness of the problem in the range of lattice
spacings used in contemporary simulations and propose a method to give more
reliable error estimates.
}

\FullConference{The XXVII International Symposium on Lattice Field Theory\\
		 July 26-31, 2009\\
		 Peking University, Beijing, China}

\begin{document}

%%%%%%%%%%%%%%%%%%%%%%%%%%%%%%%%%%%%%%%%%%%%%%%%%%%%%%%%%%%%%%%%%%%%%%%%%%%%%%%%
\section{Introduction}
%%%%%%%%%%%%%%%%%%%%%%%%%%%%%%%%%%%%%%%%%%%%%%%%%%%%%%%%%%%%%%%%%%%%%%%%%%%%%%%%
Simulations of QCD on the lattice seem to have reached the stage at which all systematic
errors can be controlled in a well defined way. This includes control over the
continuum limit $a\to0$ for which calculations at a series of decreasing lattice
spacings $a$ have to be performed. Approaching the critical point at $a=0$ using
contemporary algorithms, however, one encounters severe critical slowing down.
As we will show in the following, it is most prominent in the topological
charge, whose auto-correlation time increases dramatically as the lattice
spacing drops below $a\approx0.08$fm. Critical slowing down is a very general
problem in Markov chain Monte Carlo simulations and also the slowing down of
topological modes has been observed before. Recently, Del Debbio, Manca and
Vicari\cite{DelDebbio:2004xh} studied the issue in the CP$^{(N-1)}$ model and
found evidence for the auto-correlation times rising exponentially with the
correlation length. Also in Yang-Mills theory\cite{DelDebbio:2002xa} and 
QCD\cite{Bernard:2003gq}  large auto-correlations associated with topology
have been observed.

Our interest in the subject was spurred by the observation of a significant
slowing down of the topological charge in our simulations with two flavors of
dynamical Wilson fermions in the context of the CLS
effort\footnote{https://twiki.cern.ch/twiki/bin/view/CLS}. We are particularly
interested in lattice spacings which are fine enough to reliably simulate a
relativistic charm quark for which we  estimate $a\approx0.04$fm to be
necessary\cite{vonHippel:2008pc}.

%%%%%%%%%%%%%%%%%%%%%%%%%%%%%%%%%%%%%%%%%%%%%%%%%%%%%%%%%%%%%%%%%%%%%%%%%%%%%%%%
\section{Dynamical simulations}
%%%%%%%%%%%%%%%%%%%%%%%%%%%%%%%%%%%%%%%%%%%%%%%%%%%%%%%%%%%%%%%%%%%%%%%%%%%%%%%%
To illustrate the problem, let us start with the dynamical simulations of
$N_f=2$ degenerate flavors of non-perturbatively improved Wilson quarks and
Wilson gauge action. We use three values of the coupling constant, $\beta=5.3$,
5.5 and 5.7, which correspond to roughly $a\approx 0.08$fm, 0.06fm and 0.04fm
respectively~\cite{DelDebbio:2006cn,wittig,DellaMorte:2007sb}. For each coupling constant, several
values of the sea quark mass are simulated 
with the DD-HMC algorithm~\cite{Luscher:2005rx}, which uses a block decomposition
of the lattice to separate the infrared from the short distance modes. As
a consequence, only gauge links within the blocks are
updated (``are active''). 
We will show in the pure gauge study that this has no noticeable effect on the
problems discussed in this section beyond rescaling by the fraction of active
links.

Each of the ensembles was generated with trajectories of length $\tau=0.5$ and
after each trajectory, the lattice is shifted randomly to change the active
links. On the saved configurations we measure the topological charge
using three times HYP smeared gauge fields~\cite{Hasenfratz:2001hp} from which we
constructed the clover field strength tensor $F_{\mu \nu}$. Then the charge $Q$
is given by 
\begin{equation}
\label{eq:Q}
Q=\frac{1}{16 \pi^2} a^4 \sum_x \,  {\rm tr}\left [ F(x) \tilde F (x)\right] \ .
\end{equation}
Although this definition of the charge does in general not give an integer and
is therefore not particularly useful for the computation of physical
observables, it is known to be well correlated with, e.g. the definition of the
charge via the number of zero modes of a chiral Dirac operator. In the end, the
precise interpretation does not matter, since a slow evolution shows a
problem of the algorithm. Also note that since $Q$ in Eq.~\ref{eq:Q} is parity
odd, its expectation value vanishes.

\begin{figure}
\includegraphics[angle=-90,width=0.33\textwidth]{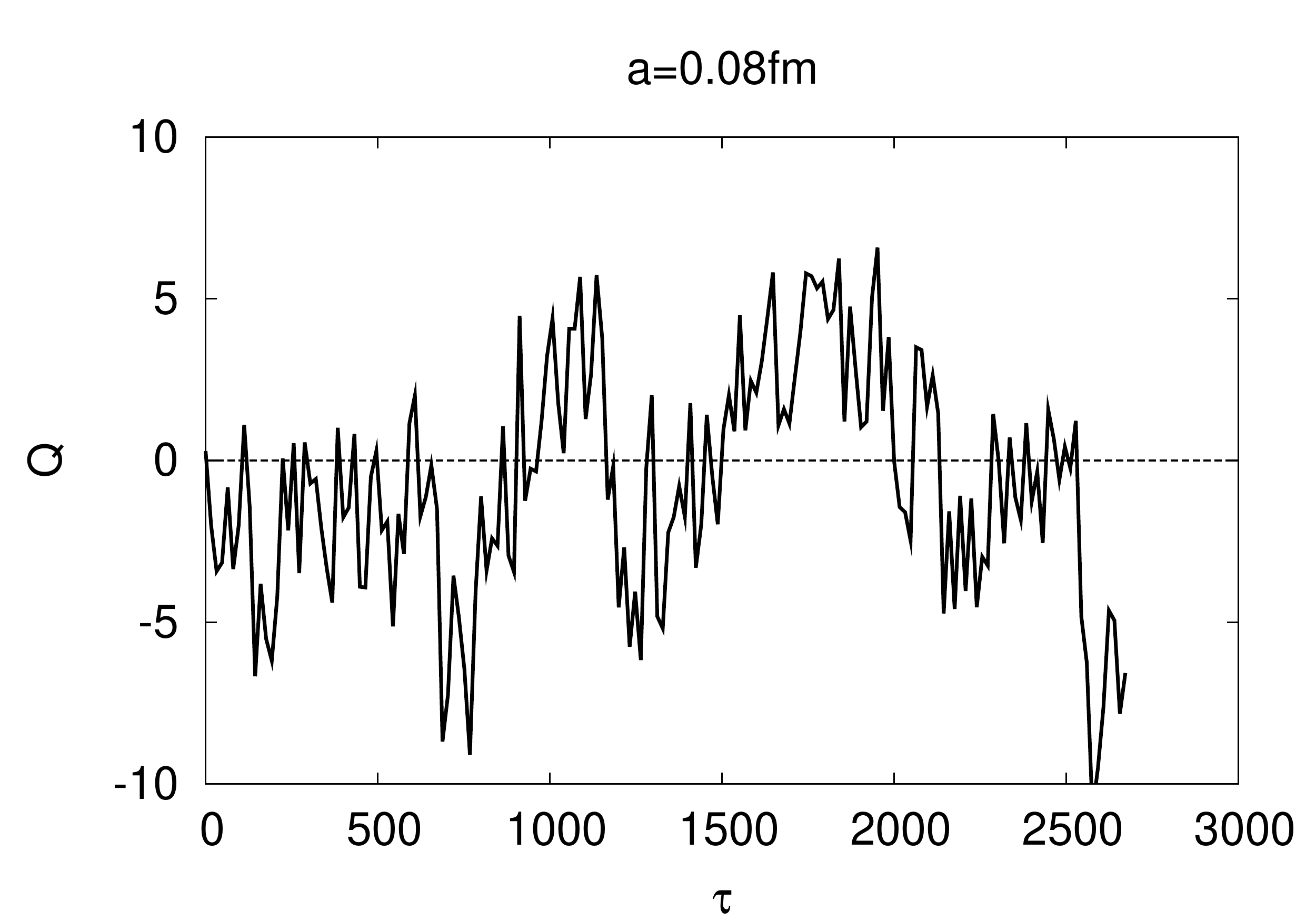}
\includegraphics[angle=-90,width=0.33\textwidth]{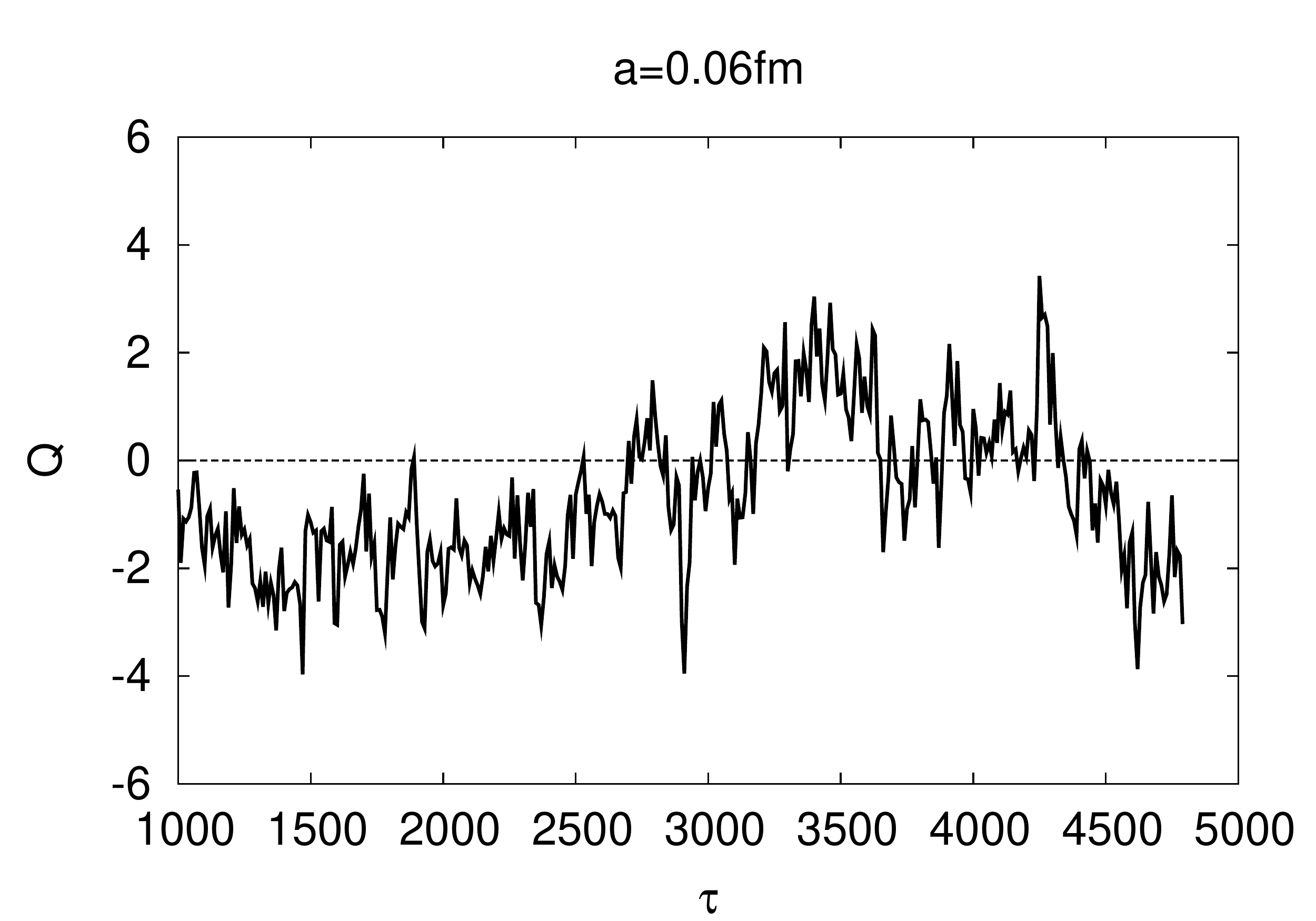}
\includegraphics[angle=-90,width=0.33\textwidth]{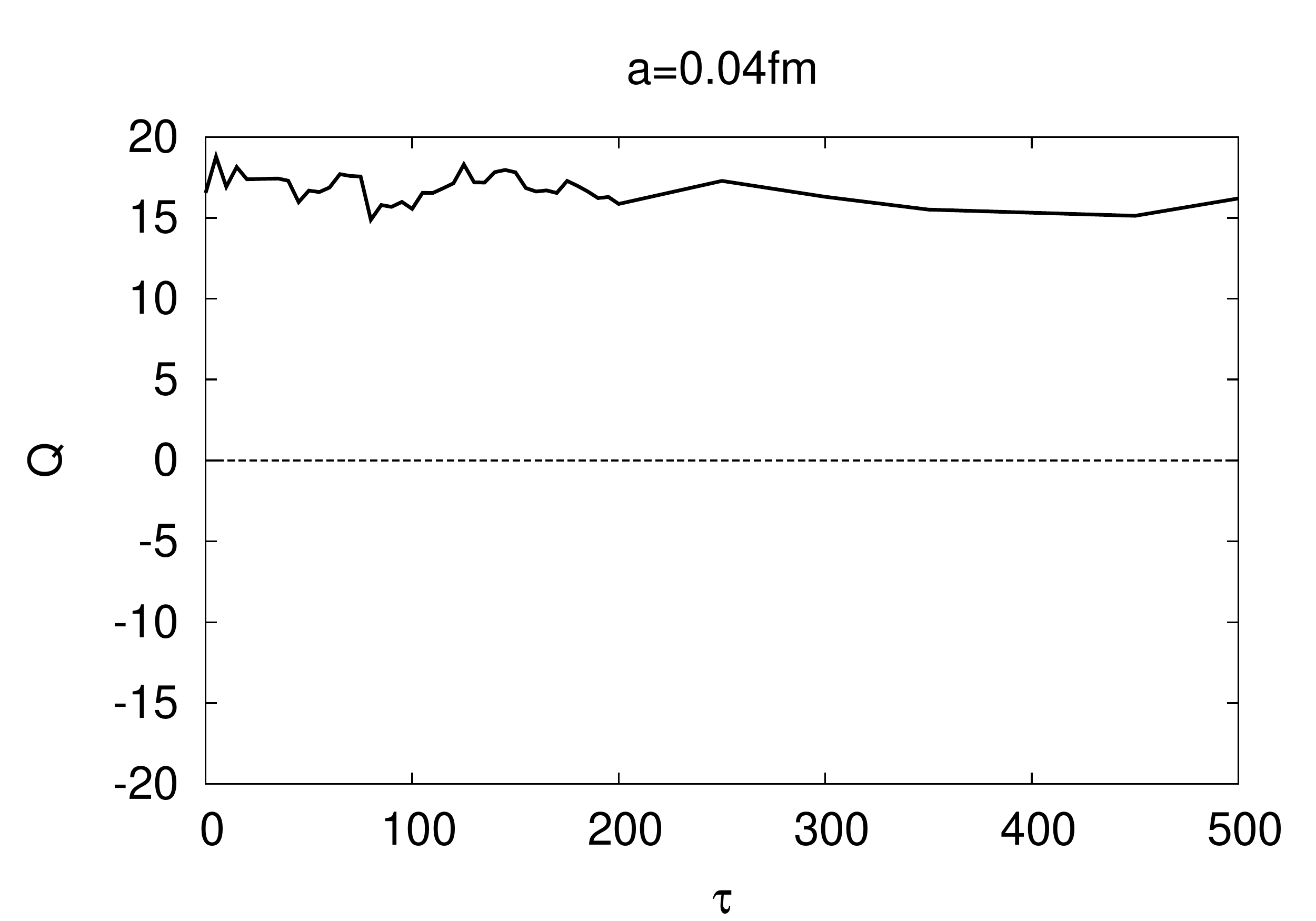}
\caption{\label{fig:1}Examples of the Monte Carlo time history of the topological 
charge. The data shown in the left plot is from a run on a $64\times32^3$ lattice
at $\beta=5.3$, $m_\pi\approx360$MeV. The central plot shows a $64\times32^3$ 
lattice at $\beta=5.5$, $m_\pi\approx460$MeV run, the right plot
a $128\times64^3$ lattice at $\beta=5.7$, $m_\pi\approx480$MeV.}
\end{figure}

Examples of the Monte Carlo time history for the three values of the coupling
constant are given in Fig.~\ref{fig:1}. At the coarsest lattice spacing
$a\approx0.08$fm, the run with roughly 5000 trajectories shows quite long
correlations, but a quite decent sampling of all topological sectors. At the
intermediate lattice spacing $a\approx0.06$fm, the auto-correlations get worse
and for the finest lattice spacing, even 1000 trajectories of length 0.5 do not
manage to move the charge considerably. The qualitative picture does not depend
on the value of the sea quark mass. What matters most is the value of the gauge coupling constant.
However, due to the  high cost of the dynamical simulations our time histories 
are too short to make a
definitive statement about the auto-correlation time of the topological charge.

%%%%%%%%%%%%%%%%%%%%%%%%%%%%%%%%%%%%%%%%%%%%%%%%%%%%%%%%%%%%%%%%%%%%%%%%%%%%%%%%
\section{Pure gauge study}
%%%%%%%%%%%%%%%%%%%%%%%%%%%%%%%%%%%%%%%%%%%%%%%%%%%%%%%%%%%%%%%%%%%%%%%%%%%%%%%%
Since  
the sea quark mass seems to have little
effect, we now study the problem in pure gauge theory. We measure the
auto-correlations of the topological charge and of Wilson loops in simulations
with the same DD-HMC algorithm as in the dynamical runs. As a comparison serve
some runs with the standard HMC algorithm, to check for effects of the
block-decomposition, and with hybrid over-relaxation (HOR) in order to see
whether this is a particular problem of the molecular dynamics based algorithms.

We use Wilson gauge action and start on a $24^4$ lattice at $\beta=6.0$ and
match lattices at constant physical volume using $r_0$ as a scale. With
$r_0=0.49$fm, the lattices of size  $L/a=16$, 24, 32 and 48 have therefore a
lattice spacing of about 0.14fm, 0.09fm, 0.07fm and 0.046fm, 
respectively~\cite{Necco:2001xg}.

\begin{figure}
\includegraphics[angle=-90,width=0.49\textwidth]{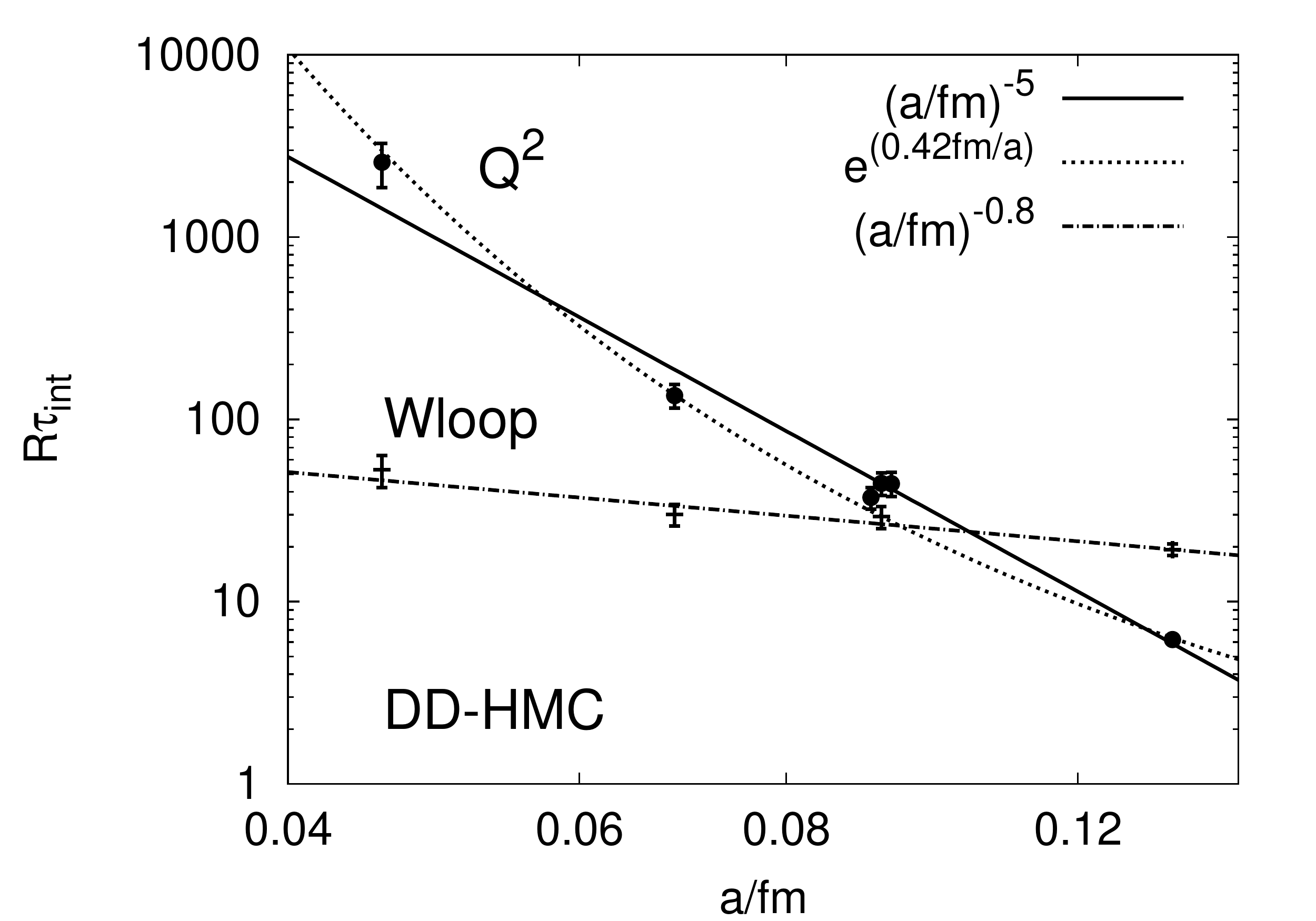}
\includegraphics[angle=-90,width=0.49\textwidth]{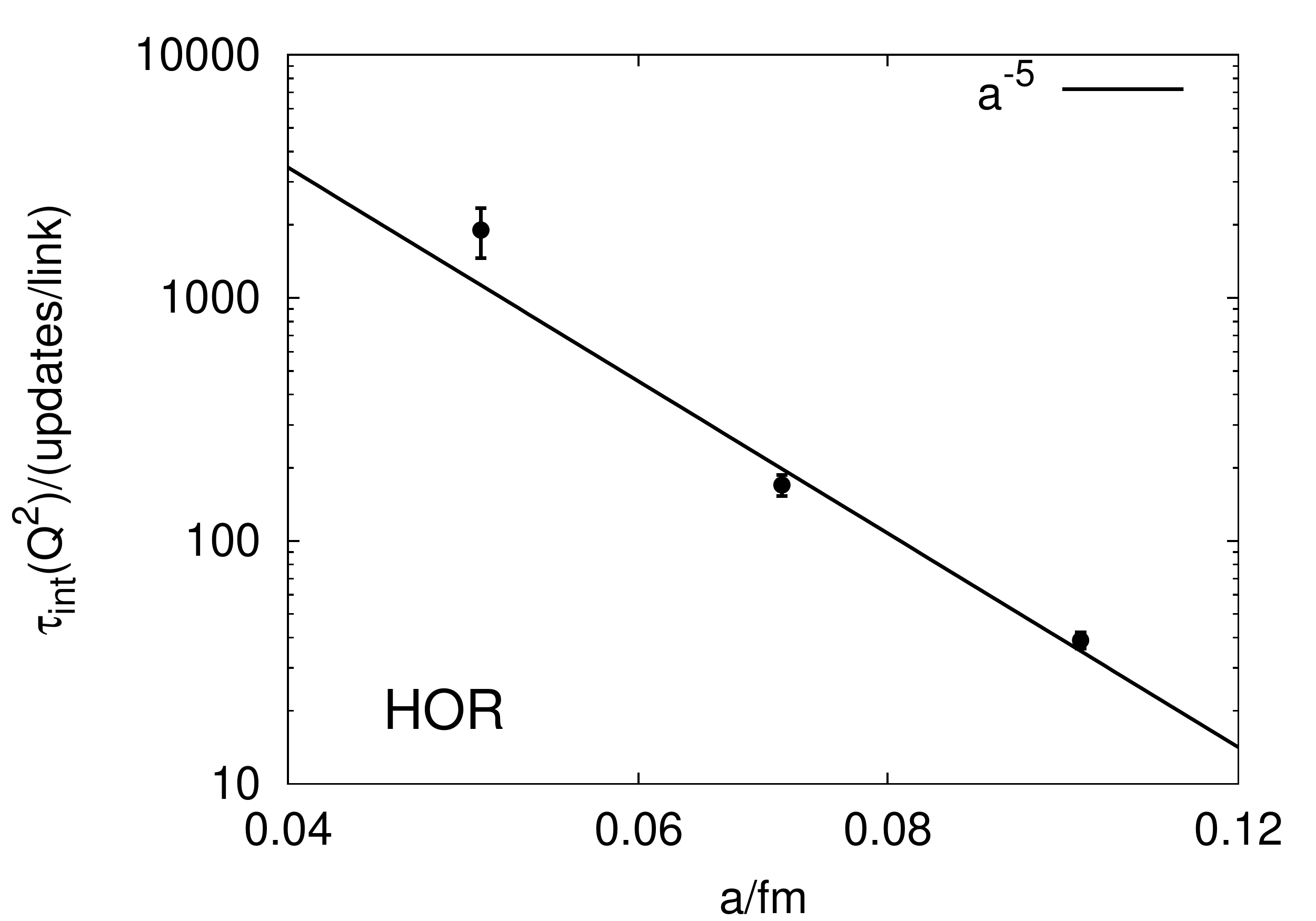}
\vspace*{-3mm}
\caption{\label{fig:crit}{\bf Left:} Auto-correlation time of $Q^2$
and the $(0.5{\rm fm})\times(0.5{\rm fm})$ square Wilson loop
as a function of the lattice spacing using the DD-HMC algorithm.
The two curves are possible functional forms which describe the data. 
At $a\approx0.09$fm we show values of $\tauint(Q^2)$ for several domain
decompositions.
{\bf Right:} Same plot, but
using the HOR algorithm on lattices of size $(4.8{\rm fm})\times(2.4 {\rm fm})^3$. 
The scales of the y-axis of the two plots are
unrelated, the line in the right plot is just to guide the eye.}
\end{figure}

With the DD-HMC algorithm and a fixed trajectory length of $\tau=4$, we measure
the auto-correlation time of $Q^2$ as a function of the lattice spacing. In the next
section we discuss the method used to determine $\tau_{\rm int}$ and
why we use $Q^2$ and not $Q$. Here, and in all other
plots, we multiply the auto-correlation times by the fraction of active links.
In pure gauge theory, also the cost of a trajectory scales with this ratio.
The result is shown on the left of
Fig.~\ref{fig:crit}. We observe a steep rise towards the continuum limit, which
is roughly compatible with $\tauint\propto a^{-5}$, but also the exponential
hypothesis of Ref.~\cite{DelDebbio:2004xh} can describe the data. Our precision
and our range in $a$ are
not enough to distinguish between the two scenarios. 

The problem is not limited to molecular dynamics based algorithms only. On the
right of Fig.~\ref{fig:crit} we show data by M.~L\"uscher and F.~Palombi using
sweeps combined of  heatbath and over-relaxation steps. The rise of the
auto-correlation time is about as steep as for the DD-HMC algorithm. This
indicates a universal problem, because the two algorithms are rather
different.

\begin{figure}
\begin{center}
\includegraphics[angle=-90,width=0.45\textwidth]{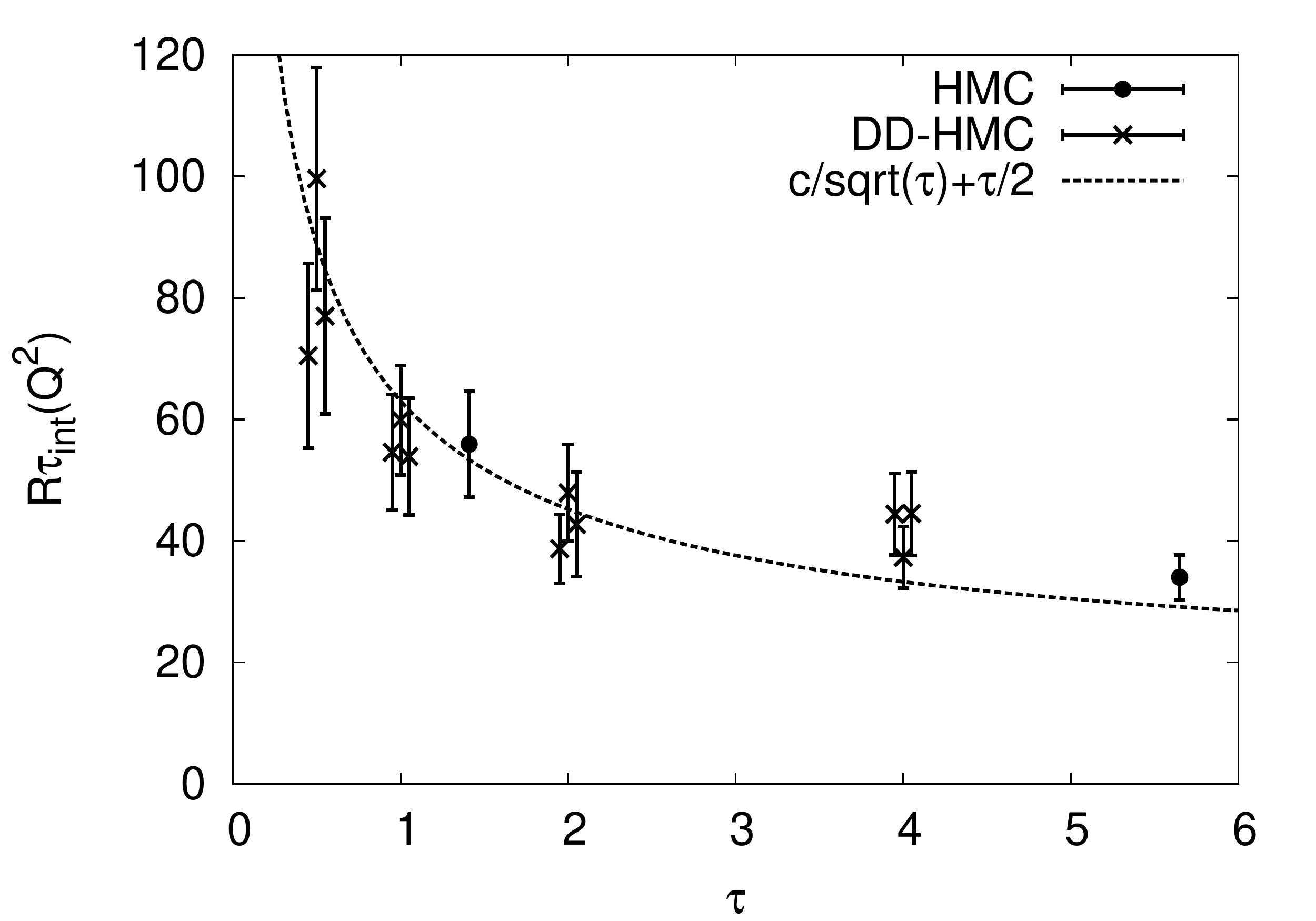}
\includegraphics[angle=-90,width=0.45\textwidth]{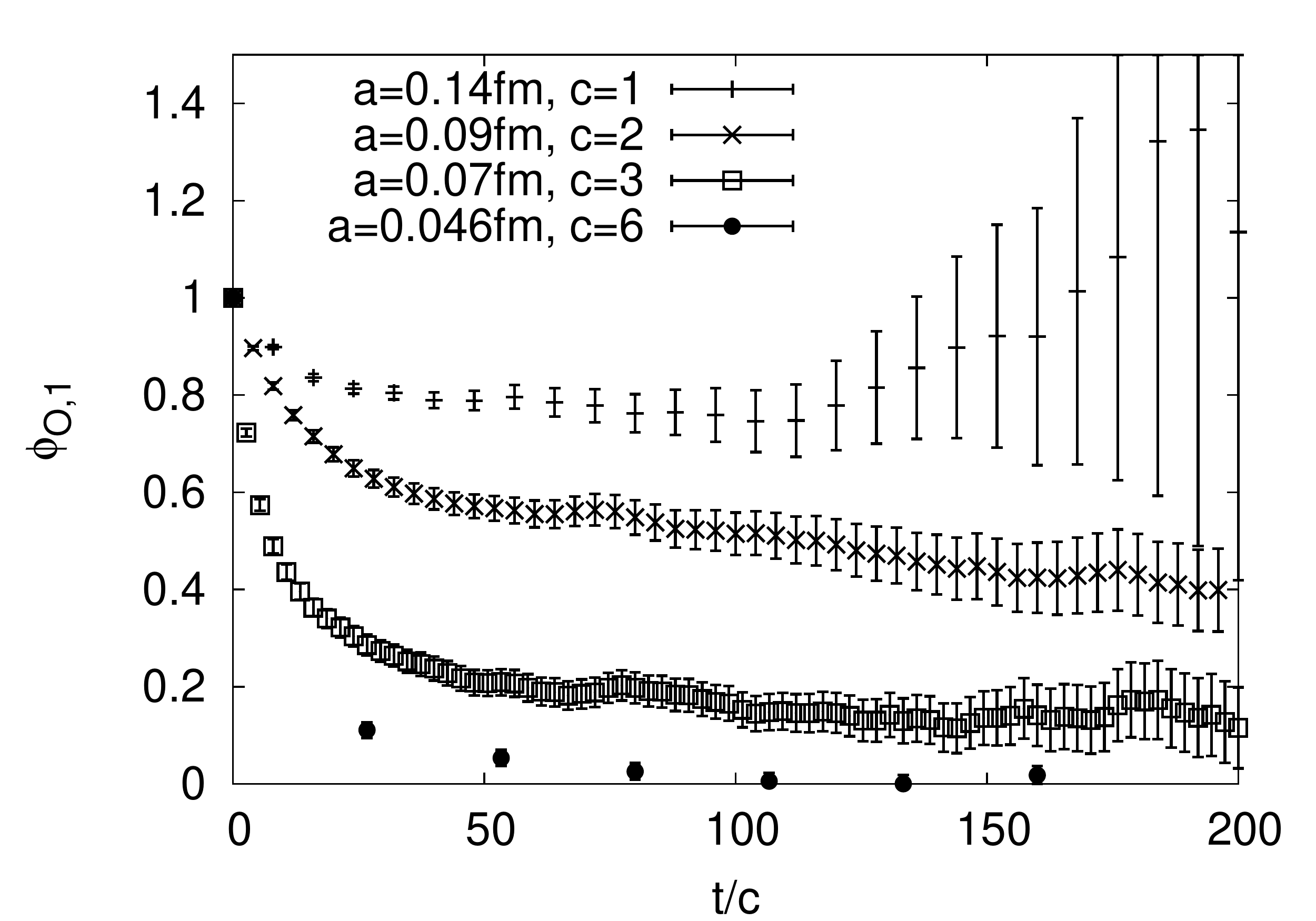}
\vspace*{-4mm}
\end{center}
\caption{
\label{fig:tau} {\bf Left:} Dependence of the AC time of $Q^2$ on the
trajectory length for the HMC and DD-HMC algorithm on the $24^4$ lattices. 
The units of $\tauint$ are
such that the differences in cost due to the inactive links has been scaled out.
The values for several block decompositions at the same $\tau$ slightly shifted
for better visibility, $6^4$, $6^2\times12^2$ and $12^4$ from left to right. 
{\bf Right:} Estimator $\varphi_{O,1}^{\rm eff}=\Gamma_O(t)/\Gamma_O(0) 
[\Gamma_{11}(t/2)/\Gamma_{11}(t)]^2$ for $\varphi_{O,1}$ of 
the (0.5fm)$^2$ Wilson loop.
}
\end{figure}

Since the auto-correlation times are much larger than 0.5, the minimum value,
a longer trajectory length might have a beneficial effect on the performance of
the algorithm. The random walk is then composed of longer, more directed 
pieces of evolution. An example for this has already been demonstrated in the
Schr\"odinger functional~\cite{Meyer:2006ty}. In Fig.~\ref{fig:tau} on the left
we show the result for $\beta=6$. We observe an
improvement---in total cost of the simulation---which is roughly compatible with
the expected $1/\sqrt{\tau}$ behavior. This gain, however, is far too small to
solve the problem. In the same 
plot we put numbers from the DD-HMC algorithm with several
block decompositions ($6^4$, $6^2\times12^2$ and $12^4$) and the HMC algorithm
without domain decomposition. All these values are compatible with another. The 
simple rescaling by the fraction of active links accounts for all the differences
in the auto-correlation times within errors.

The question arises in how far other observables are affected. This can be only
answered on a case by case basis. Here we study rectangular Wilson loops, both,
constructed from thin links and from once HYP smeared links, in particular the
one with the longest auto-correlation time, the smeared (0.5fm)$\times$(0.5fm)
square loop. Fig.~\ref{fig:tau} shows the result. On coarse lattices $\tauint$
of the loop and the charge are comparable, however, the loops are much less
affected by critical slowing down as one approaches the continuum. 
This can be
interpreted as a dynamical decoupling of the topological modes from the 
Wilson loops as the continuum limit is approached.

%%%%%%%%%%%%%%%%%%%%%%%%%%%%%%%%%%%%%%%%%%%%%%%%%%%%%%%%%%%%%%%%%%%%%%%%%%%%%%%%
\section{Stabilized error analysis and dynamical (de)coupling}
%%%%%%%%%%%%%%%%%%%%%%%%%%%%%%%%%%%%%%%%%%%%%%%%%%%%%%%%%%%%%%%%%%%%%%%%%%%%%%%%
The previous sections present the description of a grave problem for lattice
simulations. The topological charge moves
very slowly in the simulations. We have already seen that other observables
may be affected less, because they are only weakly coupled to this ``mode''.
Here we want to discuss what such a decoupling means and how error estimates 
can be stabilized in this difficult situation.

We consider an observable $O=F(\bar A)$, which is 
a function of the mean $\bar A_\alpha = \langle A_\alpha \rangle$ 
of primary observables $A_\alpha=A_\alpha^*$ 
averaged over $N$ subsequent measurements in the Markov chain.
The error, $\delta O$ is given by
$ (\delta O)^2 = {\rm var}(O) \frac{2 \tauint(O)}{N} $,
where the variance is a property of the theory and the auto-correlation
time a  property of the algorithm. It is given~\cite{Wolff:2003sm} 
by the summed auto-correlation function $\Gamma(t)$
\begin{equation}
\label{eq:tauint}
\begin{split}
\tauint(O)& =\frac{1}{2}+\sum_{t=1}^\infty \frac{\Gamma_{O}(t)}{\Gamma_{O}(0)} \, , \\
 \Gamma_{O}(t) 
 &= \sum_{\alpha\beta}\,F_\alpha F_\beta\, \Gamma_{\alpha\beta}(t)\, , \qquad
 \Gamma_{\alpha\beta}(t) 
 = \langle (A_\alpha(t) - \bar A_\alpha  ) (A_\beta(0) - \bar A_\beta  ) \rangle \ ,
\end{split}
\end{equation}
where $t$ is the separation in Monte Carlo time and 
$F_\alpha=\partial F/\partial\, \bar A_\alpha$. 

The function $\Gamma$ can be expressed in terms of the 
transition matrix of the Markov chain, $M(q',q)$, which gives the transition probability
for a change from a state $q$ to a state $q'$. Denoting the equilibrium
probability distribution by $W(q)$, and assuming the algorithm to satisfy
detailed balance  means that the matrix 
$T(q',q)=(W(q'))^{-1/2}\,M(q',q)(W(q))^{1/2}\,$ is symmetric. Its eigenfunctions
are labeled $\chi_n(q)$ and its eigenvalues $\lambda_n$. The 
autocorrelation function can then be written as
\begin{equation}
    \label{e:modes}
    \Gamma_{\alpha\beta}(t) = \sum_n (\lambda_n)^t \eta_{n,\alpha}^* \eta_{n,\beta} 
     = \sum_n [{\rm sign}(\lambda_n)]^t \ {\exp(-t/\tau_n) } \ \eta_{n,\alpha}^* \eta_{n,\beta} \ ,
\end{equation}
where $\lambda_n<1$ and $\tau_{n+1}\leq\tau_n=-1/\ln|\lambda_n|$ and 
\begin{equation}
  \eta_{n,\alpha} = (\chi_n, W^{1/2} A_\alpha) = \int {\rm d} q\,\chi_n^*(q)\,W^{1/2}(q) A_\alpha(q)\,.
\end{equation}
In general all modes $n$ contribute to the sum in
Eq.~(\ref{e:modes}) and slow modes have $\tau_n\gg1$ producing a long tail
in $\Gamma(t)$. This 
is typically hard to
estimate: one quickly reaches the point from which the uncertainty 
of $\Gamma(t)$ is larger
than the signal. Therefore one truncates the sum in Eq.~\ref{eq:tauint} at some
time $w$. This, however, introduces a bias which can be sizeable if the tail is
very long.

There can be  
selection rules imposed by common symmetries of the theory (i.e. $W(q)$) and
the algorithm (i.e. $T(q',q)$). Parity is such a case relevant for
us: $Q$ couples only to parity odd modes, while $Q^2$ and other parity
even observables couple only to even modes.
Since in our lattice QCD discretization parity is an exact symmetry,
parity odd observables vanish and
we only need to consider even ones. Hence
only those $n$,
corresponding to parity even $\chi_n(q)$ contribute in Eq.~(\ref{e:modes}). 

\begin{figure}
\begin{center}
\includegraphics[angle=-90,width=0.45\textwidth]{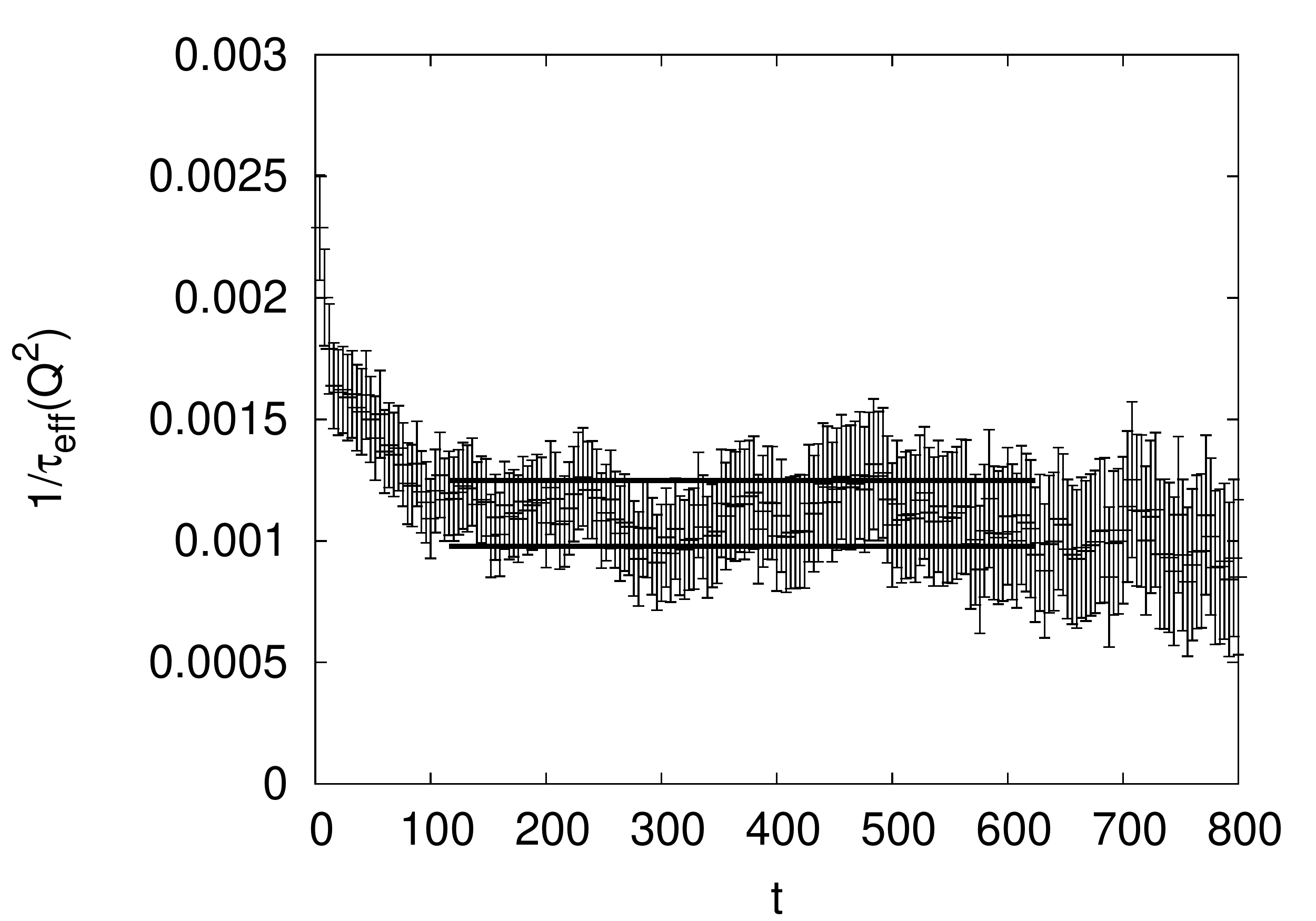}
\includegraphics[angle=-90,width=0.45\textwidth]{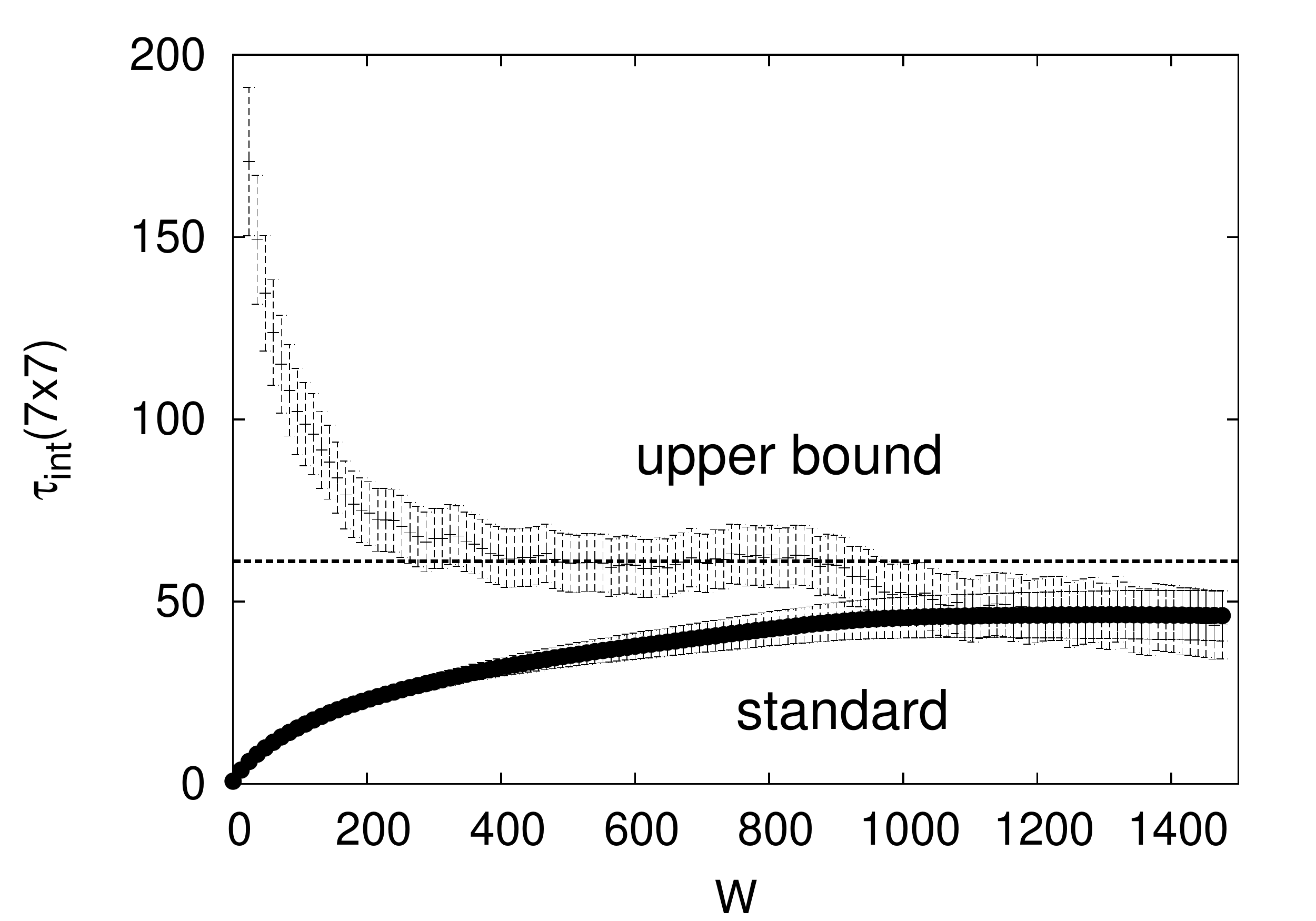}
\vspace*{-4mm}
\end{center}
\caption{\label{fig:tauexp}{\bf Left:} Extracting the exponential auto-correlation time
from an effective mass plot of $\Gamma(t)$ of $Q^2$ for our $32^4$, $\tau=1$ run with
the DD-HMC algorithm.
{\bf Right:} Estimate of $\tauint$ of the 0.5fm square
Wilson loop as a function of the summation window according to
Eq.~{\protect\ref{e:uppb}}.
}
\end{figure}

In order to learn about the slowest mode in a simulation, we can then investigate 
an ``effective mass'' plot.\footnote{Experience shows that oscillating terms (${\rm sign}(\lambda_n)=-1$) are usually 
irrelevant at large $t$.} Indeed we find a
long plateau in $1/\tau_{\rm eff} = \log(\Gamma_{11}(t)/\Gamma_{11}(t-1))$ for 
$A_1=Q^2$ in Fig.~\ref{fig:tauexp}. It provides a good estimate for 
$1/\tau_{\rm exp}=-\log(\lambda_1)$. 

If $\lambda_1$ has been determined well, it can be inserted into 
the (for even $w$) exact inequality 
\begin{equation}
\tauint \leq
\frac{1}{2} + \sum_{t=1}^{w-1} \frac{\Gamma_O(t)}{\Gamma_O(0)} 
             + \frac{1}{ 1-\lambda_1}  \frac{\Gamma_O(w)}{\Gamma_O(0)} \ 
\approx 
\frac{1}{2} + \sum_{t=1}^{w-1} \frac{\Gamma_O(t)}{\Gamma_O(0)} 
             + \tau_{\rm exp}  \frac{\Gamma_O(w)}{\Gamma_O(0)} 
\label{e:uppb} 
\end{equation}
to obtain an upper bound for the error. 
\footnote{Actually, the DD-HMC
algorithm does not fulfill detailed balance, because of the directed
block shifts, and the formulae do not apply.
However, we expect this to have little effect on the large time behavior
of $\Gamma(t)$.
}

In a practical procedure we first select a set of observables
(plaquette and $Q^2$ after various smearing levels), determine
approximate plateaux in their ``effective mass'' $1/\tau_{\rm eff}$ and take the largest one
as the estimate for $\tau_{\rm exp}$ in Eq.~(\ref{e:uppb}). Note that if the tail of the 
auto correlation function is dominated by just the slowest mode or
by modes with eigenvalues close to it, the upper bound will 
(almost) be saturated. 
That such an upper bound is not overly conservative is demonstrated in
Fig.~\ref{fig:tauexp} on the right, on the $0.07$fm lattice where the
$\tau_{\rm eff}$ is taken from the topological charge and the observable is the
$7\times7$ Wilson loop. We find that generically the upper bound converges
faster as a function of the window size $w$ than the more standard 
estimate $\tauint= 1/2+\sum_{t=1}^w \frac{\Gamma_O(t)}{\Gamma_O(0)}$. 
We therefore advocate its use in difficult situations, e.g. in HMC
simulations of QCD.

Let us now return to the decoupling of some observables from the 
critical slowing down. Since in the representation Eq.(\ref{e:modes})
all modes $n$ contribute, a decoupling from the 
slowest mode means $|\sum_\alpha F_\alpha\eta_{1,\alpha}|^2/\Gamma_O(0) = \varphi_{O,1} \ll 1$. 
Combining
the autocorrelation functions of the Wilson loops and those
of $Q^2$, one can easily form estimators which converge to 
$\varphi_{O,1}$ for large $t$. Even if we observed this convergence to be
rather slow, we found that the estimators for $\varphi_{O,1}$
become tiny at small lattice spacing, see Fig.~\ref{fig:tau} (right).
This holds for Wilson loops and Creutz ratios.

%%%%%%%%%%%%%%%%%%%%%%%%%%%%%%%%%%%%%%%%%%%%%%%%%%%%%%%%%%%%%%%%%%%%%%%%%%%%%%%%
\section{Summary}
%%%%%%%%%%%%%%%%%%%%%%%%%%%%%%%%%%%%%%%%%%%%%%%%%%%%%%%%%%%%%%%%%%%%%%%%%%%%%%%%
In this contribution we have studied the critical slowing down of lattice
simulations using the HMC and DD-HMC algorithms. We demonstrated a steep
increase of the auto-correlation time of the squared topological charge 
in pure gauge
simulations which match our observations with dynamical fermions. Tuning of the
parameters of the algorithm can improve the situation a bit, but probably more
dramatic changes like the one proposed in \cite{Luscher:2009eq} will be needed
to solve the problem.

In the second part, we studied the impact of such slowly moving modes on the
estimation of the errors in Monte Carlo simulations. We proposed a method to
give upper bounds on the contribution from the tail of the auto-correlation
function, which is normally neglected. The resulting estimates are neither overly
conservative nor impractical and therefore are a safer way to determine the
errors.

%%%%%%%%%%%%%%%%%%%%%%%%%%%%%%%%%%%%%%%%%%%%%%%%%%%%%%%%%%%%%%%%%%%%%%%%%%%%%%%%
\acknowledgments
%%%%%%%%%%%%%%%%%%%%%%%%%%%%%%%%%%%%%%%%%%%%%%%%%%%%%%%%%%%%%%%%%%%%%%%%%%%%%%%%
%\footnotesize
Work supported in part by the SFB/TR~9 of the Deutsche Forschungsgemeinschaft
and by the European Community
through EU Contract No.~MRTN-CT-2006-035482, ``FLAVIAnet''.
Our simulations are performed on BlueGene and PC-clusters of
the John von Neumann Institute for Computing
at FZ J\"ulich, the HLRN in Berlin, the Universities of Mainz, Rome La Sapienza,
Valencia-IFIC and at CERN, as well as on the IBM MareNostrum
at the Barcelona Supercomputing Center.
We thankfully acknowledge the computer resources and support provided
these institutions.
We thank M.~L\"uscher and
F.~Palombi for very interesting discussions and providing us with unpublished data
and C. DeTar for a very useful email exchange.

\end{document}